\documentclass{article}

\usepackage{todonotes}

\usepackage{stackengine}

\usepackage{multirow}
\usepackage{url}

\usepackage[margin=1in]{geometry}

\usepackage{ulem}

\usepackage[colorlinks = true,
            linkcolor = blue,
            urlcolor  = blue,
            citecolor = blue,
            anchorcolor = blue]{hyperref} 

\usepackage[parfill]{parskip}
\usepackage{titlesec}

\usepackage{fourier-orns}

\usepackage{pifont}
\usepackage{amssymb}

\usepackage{leipzig}

\usepackage{caption} 
\usepackage{booktabs} 

\usepackage{ot-tableau}

\usepackage{color, soul}

\usepackage[linguistics]{forest}

\forestset{
  nice nodes/.style={
    for tree={
      inner sep=0pt,
      fit=band,
    },
  },
  default preamble=nice nodes,
}

\usepackage{gb4e}

\makeatletter         
\def\@maketitle{   
\noindent {\Large \bfseries \color{black} \@title}  \\ \hrule \noindent \@author \\ \@date  
}

\usepackage{tikz}

\RequirePackage{leipzig}

\RequirePackage{xspace}
\xspaceaddexceptions{]\}}

\newleipzig{hab}{hab}{habitual}
\newleipzig{rem}{rem}{remote}
\newleipzig{sm}{sm}{subject marker}
\newleipzig{t}{t}{tense}
\newleipzig{aa}{aa}{anti-agreement}
\newleipzig{pron}{pron}{pronoun}
\newleipzig{rec}{rec}{recent}
\newleipzig{om}{om}{object marker}
\newleipzig{asp}{asp}{aspect}
\newleipzig{lk}{lk}{linker}
\newleipzig{pcl}{pcl}{particle}
\newleipzig{stat}{stat}{stative}
\newleipzig{ints}{ints}{intensive}
\newleipzig{ascl}{ascl}{assertive subject clitic}
\newleipzig{nascl}{nascl}{non-assertive subject clitic}
\newleipzig{ta}{ta}{tense and/or aspect}
\newleipzig{assoc}{assoc}{associative marker}
\newleipzig{hon}{hon}{honorific}
\newleipzig{sa}{sa}{subject agreement}
\newleipzig{conj}{conj}{conjunction}
\newleipzig{expl}{expl}{expletive}
\newleipzig{rcm}{rcm}{reciprocal marker}
\newleipzig{pers}{pers}{persistive}

\usepackage{phonrule}

\usepackage{textgreek}

\usepackage{metalogo}

\title{NDN-TR70 - Utilizing NDN-DPDK for Kubernetes Genomics Data Lake}
\author{\textbf{Sankalpa Timilsina, Justin Presley, David Reddick, Susmit Shannigrahi}, \\\textit{Tennessee Tech}, \\\textbf{Xusheng Ai, Coleman Mcknight, Alex Feltus}\\\textit{ Clemson University}}

\date{\today} 

\begin{document}

\maketitle

\begin{abstract}

\noindent\textbf{As the growth of genomics samples rapidly expands due to increased access to high resolution DNA sequencing technology, the need for a scalable platform to aggregate dispersed datasets enable easy access to the vast wealth of DNA sequences available is paramount. In this work, we introduce and demonstrate a novel way to use Named Data Networking (NDN) in conjunction with a Kubernetes cluster to design a flexible and scalable genomics Data Lake in the cloud.  In addition, the use of the NDN Data Plane Development Kit (DPDK) provides an efficient and accessible distribution of the datasets to researchers anywhere. This report will explain the need to deploy a Data Lake for genomics data, what is necessary to deploy successfully, and detailed instructions to replicate the proposed design. Finally, this technical report outlines future enhancement options for further improvements.}

\end{abstract}

\section{Introduction}

A Data Lake is a large-scale data repository for raw/intermediate file storage for data analytics. Why create a Data Lake for genomics data? Since scientists constructed the reference human DNA genome over two decades ago, there has been an explosion of genomes sequenced from thousands of organisms across the tree of life. Each genome sequence is the blueprint for that species and serves as a scaffold for organizing new knowledge on the flow of information from DNA to a trait. While interesting from a general biology perspective, the applications of merging the genome with other genomics datasets have practical implications, including applications in medicine, agriculture, bioenergy, and other fields.

A barrier to using genomics datasets is that they are often difficult to find and access (download) from multiple public repositories and merge with other public or with local data so that they may become co-analyzed (interoperable). This makes data un-FAIR (Findable, Accessible, Interoperable, Reuseable; \cite{FAIR}). Here we propose a FAIR Data Lake where genomics datasets from different sources are named and co-published without the end-user needing to know how to transfer data to the Data Lake. Datasets are then pulled from the data lake using an intuitive naming convention based on Named Data Networking (NDN; \cite{ndnwebsite}). An overview of genomics data naming can be found in \cite{ogle2021named}.

Since genomics datasets are massive (gigabyte to terabyte-scale), the Data Lake must be a flexible, optimized, and scalable system. We believe that clouds (public or private) are the scalable solution, so we are building a Data Lake framework for the cloud. Specifically, we will build the data lake using containers that run on a Kubernetes (K8s; \cite{Kubernetesio})-based system so that the Data Lake can scale dynamically as well as be deployed on multiple commercial and public clouds. One of the critical problems of location-independent deployment has been management complexity. While current data lakes can be deployed on any cloud platform, they need to be configured manually to be accessible to the users. The data also depends on the IP address and DNS name of the platform where the data lake is deployed.

We will use an open-source Data Plane Development Kit (DPDK) for network transport that allows location-independent optimization. This technical report describes the design and implementation of the NDN Data Lake for the cloud.

\section{Background}

\subsection{Named Data Networking}
Currently, networking for layer 3 of the OSI model \cite{osimodel} uses the Internet Protocol (IP) that implements the Host-based networking (TCP/IP) paradigm in which data is transferred from one host to another using a location-based address (IP address). In contrast to IP, Named Data Networking (NDN) \cite{zhang2014named} is one of the most developed and implemented designs using the Information-Centric Networking (ICN) paradigm. ICN fully removes the location requirement that was introduced with the TCP/IP paradigm and shifts the focus from location to data.

Transferring data via NDN from one host to another is accomplished via two types of packets: Interest packets and Data packets. Receiving data is as simple as expressing (sending) an Interest packet and receiving a matching Data packet. The applications decide which name(s) they are going to request. Once an Interest is expressed, NDN utilizes name-based forwarding to forward the packet toward the data source. Data is also signed and can be verified by the recipient; therefore, it can come from anywhere: a publisher, a proxy, or an in-network cache.

By using NDN, consumers and producers can utilize several benefits, such as data availability after server failures, a significant decrease in server traffic, and faster data retrieval. Instead of securing a connection, NDN secures the data, removing most connection-oriented attacks, including a Man-In-The-Middle attack. In addition, serving and replicating data across nodes is built into NDN. In NDN, the names are hierarchical, similar to the HTTP Uniform Resource Locator (URL). However, NDN names are Uniform Resource Identifiers (URIs) -  unlike URLs, they point to a piece of content and not the location of the content. Hierarchical names provide the ability to reduce in-network state as well as make discovery easier. All these properties make NDN an excellent mechanism for constructing a named genomics Content Delivery Network (CDN) -- the Genomics Data Lake.

\subsubsection{Forwarders}
In order to utilize the NDN architecture, a forwarder must be present in order to route and fulfil NDN interests properly. NDN-DPDK \cite{NDNDPDK} and NDN Forwarding Daemon (NFD) \cite{afanasyev2018nfd} are network forwarders for NDN that support Interest and Data forwarding as well as content caching in the network. This is accomplished by abstracting lower-level network transport technologies into NDN Faces, maintaining fundamental data structures such as CS, PIT, and FIB, and implementing packet processing logic\cite{NFDdevguide}. 

NDN-DPDK \cite{NDNDPDK} is a high-speed NDN forwarder developed with the Data Plane Development Kit (DPDK)\cite{dpdkwebsite}. DPDK includes data plane libraries and polling-mode network interface controller drivers for offloading TCP packet processing from the operating system kernel to user-space programs. This offloading enables better computational efficiency and packet throughput than is attainable with the kernel's interrupt-driven processing.

With NFD, high-speed forwarding is still a challenge due to variable-length named-based lookups as well as packet state updates. In this project, we choose the NDN-DPDK forwarder due to its performance advantages over NFD. While running on commodity hardware, the NDN-DPDK forwarder can reach a forwarding speed of more than 100 Gbps \cite{ndndpdkpaper}. This will be useful in transferring data between NDN data lakes and cloud deployments (such as the Pacific Research Platform (PRP) Kubernetes cluster \cite{smarr2018pacific}).

\subsection{Kubernetes}

In recent years, container technology has been gaining increasing traction. A container is a unit of software that bundles code and all dependencies required for the app to run. Containers also lend themselves well to the architectural approach where an application is separated into multiple services that rely on each other to perform the full desired function.\cite{micro}. Kubernetes \cite{Kubernetesio} is an open-source container orchestration framework that provides an ideal platform for automating containerized applications in different deployment environments. Kubernetes typically deploys containers in environments with high bandwidth and low latency network connections. The applications should spread across the service nodes with high availability, applications are always accessible by the users, and scalability, scaling application fast when more users are trying to access it. Moreover, Kubernetes has a disaster recovery mechanism that prevents users from losing any data when hardware or software failures happen to the service center. 

\subsubsection{Limitations of external inbound access.}

When working with Kubernetes, internal networking inside a cluster has default communication rules that facilitate networking between services on different hosts in a cluster. This communication is further facilitated by using cluster IPs attached to each pod, allowing for accessible internal communication but not external. While only having communication inside a Kubernetes cluster may be sufficient for some applications, others require external access. A Kubernetes service like NodePorts, NGINX, or a load balancer will be necessary to resolve this communication limitation \cite{services1}\cite{EKS}.

\subsection{Data Lake}

A Data Lake is a repository that centralizes raw data for storage from many data sources into a single data store. This data repository is designed to handle large amounts of data ranging from multiple terabytes to petabytes. The data is long-lived, unprocessed, and left in a form that other services can access and use for analytics or machine learning, depending on the application \cite{lake-aws}\cite{lake-google}\cite{lake-mircro}.

\section{Overview}

\begin{figure}[h!] 
\centering
 \includegraphics[width=0.4\textwidth]{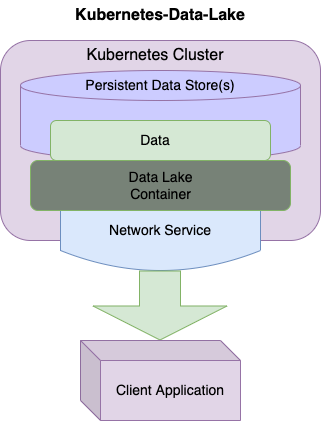}
 \caption{The figure demonstrates a high-level overview of the proposed Data Lake using Kubernetes. As shown, the data storage and docker containers are contained inside the Kubernetes cluster with networking services used to provide external access to a client application. }
\label{fig:data-lake-high}
\end{figure}
The proposed Data Lake for genomics research is built upon the highly scalable platform of Kubernetes. Inside Kubernetes, multiple functions are required to enable the successful creation and access of the novel genomics Data Lake that utilizes NDN and DPDK to discover and deliver needed samples. To assist with comprehension of the overall design that will later be explained in more detail later in this report, Figure \ref{fig:data-lake-high} provides a high-level overview of the proposal. As discussed before, one of the first and most essential components is Kubernetes. All the storage and features needed to facilitate the content delivery are run inside a Kubernetes cluster for this deployment. When working with data at an expansive scale, like what is required for modern genomics research, the ability to store large quantities of data is paramount. When working with Kubernetes, persistent data stores provide the necessary storage to hold the desired data, as shown in the figure. Once information is stored in Kubernetes, a way to access and retrieve the data is also needed. For this, docker containers are deployed on the Kubernetes cluster to provide an interface to the stored data and create the Data Lake. On these containers, NDN DPDK applications are run to give the client access to the data. With the containers providing access to the stored data, the final component with Kubernetes is the network services. Internally access inside the Kubernetes cluster is accomplished using cluster IPs; however, this only allows access from inside the cluster. In a real-world deployment, access to the data externally over the Internet is required and accomplished using Kubernetes network services like NodePorts. While these components outline what is needed to create a Data Lake on Kubernetes, accessing the data is also essential for practical use. While the client application will vary greatly depending on the end user's requirements, options like Nextflow are an option to provide integration with an existing genomic research workflow with the outlined Data Lake.
\section{Design}

\subsection{Requirements}

The NDN-DPDK virtualized cloud architecture underlying the Data Lake aims to combine the best features of the NDN-DPDK network architecture with the best features of virtualization. NDN-DPDK utilizes the full software and network stack of the machine running it down to the network card and kernel. While this design serves to maximize the performance of the architecture, it also presents some challenges when running the NDN-DPDK software in any virtual environment, especially within containers, which by nature, share the kernel of the host operating system. Additionally, the container orchestration platform must accommodate atypical virtualization and network traffic. This section describes the requirements of a Kubernetes cluster for deploying the NDN-DPDK virtualized cloud architecture. Ideally, the future will see K8s clusters come with these features as a standard design to support Future Internet architectures.

\subsubsection{Layer 2, 3 and Memif Connections}
Our deployment runs NDN-DPDK, which uses the docker containers of the Data Lake to communicate over the data link layer to utilize the NDN data routing protocols instead of standard TCP/IP. Ideally, the interfaces within each docker container must be able to connect to interfaces on other containers using the interfaces' MAC addresses. However, this is not always a simple task, especially in Kubernetes clusters deployed over a Wide Area Network and using IP by default. Layer 2 communication between Data Lake containers may be achieved through connected Virtual Local Area Networks(VLANs) at layer 2. It is also possible to connect the containers over Layer 3, but that approach demonstrated decreased performance. NDN-DPDK deployed over Layer 2 makes the incoming traffic bypass the TCP/IP stack and is made accessible directly from the user space. With Layer 3 deployment, the incoming traffic uses the usual TCP/IP stack and is comparatively much slower. Likewise, the client outside the Data Lake can be connected to the high-speed forwarder on the Data Lake over Layer 2 or Layer 3. The forwarder is connected to the NDN-DPDK's fileserver application on the same node using the shared memory packet interface (memif). Memif provides a very high-performance memory-based interface. It functions by exchanging a shared memory over a UNIX file socket and using this memory for transmitting and receiving packets.  



\subsubsection{Cluster Networking}
Kubernetes clusters use a virtual overlay network to allow containers to communicate with one another, referred to as the “pod network” \cite{clusternetworking}. Virtualizing networks within a cluster has many benefits, such as eliminating the need for developers to coordinate which ports they are using with the other users of the cluster. This does come with drawbacks as containers must be exposed by predefined services \cite{k8sservices}, and these services expose containers in different capacities. Internal traffic is achieved by exposing containers to the virtual pod network with services called \textbf{ClusterIPs}, while the \textbf{LoadBalancer}, \textbf{HostPort}, and \textbf{ExternalName} services expose the container to the external Internet \cite{k8sservicestypes}. 

Although NDN-DPDK can use layer 3 with standard TCP/UDP protocols, the services must also allow for TCP/UDP. The only services that allow TCP/UDP traffic are ClusterIPs and HostPort, so a Kubernetes cluster hosting the NDN-DPDK virtualized cloud architecture must allow for both to be used. ClusterIPs are the default service and are supported by every K8s cluster; however, HostPorts are not. Support for HostPorts is necessary to deliver content through a port of the NDN-DPDK container exposed on a port of the host node. 





\subsection{Data Loading}

\begin{figure}[h!] 
\centering
 \includegraphics[width=1.0\textwidth]{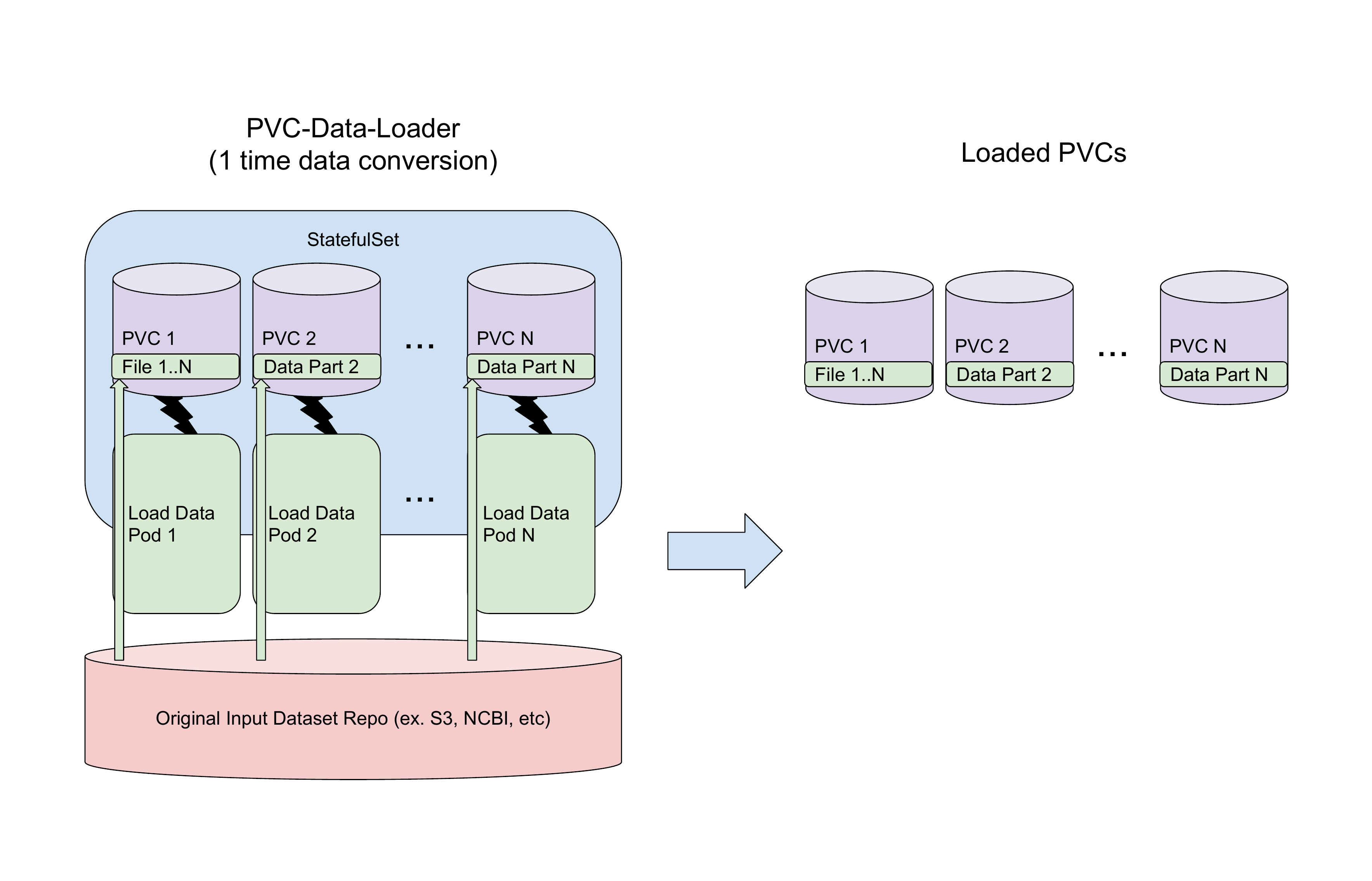}
 \caption{The cloud architecture of the PVC-Data-Loader. On the left, data is pulled from its original repository using the containerized client software of the desired protocol. The official amazon/aws-cli\cite{awscli} image is used as an example. The right side shows the loaded PVCs to be used.}
\label{fig:pvc-data-loader}
\end{figure}

To establish a Data Lake, a significant amount of attention must be paid to data movement. This of course includes the egress of content being delivered to consumer applications but also the ingress of the content to be published. How can we import massive amounts of data into a Kubernetes cluster in a sensible way? Using various application layer protocols, the data could be of any size and from any source. The data must be loaded onto each node hosting the Data Lake and be tolerant to faults on the virtual machine, host container, host node, and entire cluster. Here we describe the PVC Data Loader \cite{pvcdataloader}, a novel tool for the parallel and distributed ingress of data into multiple Persistent Volume Claims (PVCs) on a Kubernetes cluster. Figure \ref{fig:pvc-data-loader} shows this architecture. 

\subsubsection{PVC Creation}

PVCs are portions of a shared file system on a Kubernetes cluster \cite{pvcs}. PVCs provide a persistent data store for K8s users while allowing multiple independent users or groups to work without the risk of mismanagement or misuse of data. The PVC Data Loader has the ability to create a specified number of PVCs automatically, using a built-in StatefulSet utility called a Volume Claim Template \cite{volumeclaimtemplates}. This is suitable for Data Lakes located within a single region or where only one NDN-DPDK producer container is mounted to each PVC. For K8s clusters with storage classes\cite{storageclasses} and nodes in different geographical regions, it is better to create the PVCs manually beforehand. That ensures the content being published is located within close proximity in terms of network topology. For Data Lakes where multiple Producer containers are mounted to the same PVC, those PVCs must be manually created with the *Read-Write-Many* Access Mode \cite{accessmodes}.

\subsubsection{Supported Protocols}

The content that a Data Lake may publish could be located anywhere on the traditional Internet used today. Thus the need exists for the PVC Data Loader to pull data from any client application that pulls data from its original repository. Fortunately, this can be achieved by utilizing the interchangeable nature of containers within Kubernetes deployments. The PVC data loader can be used with any containerized client protocol by changing the *image* field in the helm configuration. The Kubernetes cluster will pull the specified image from the container repository that it is published from. Examples of container repositories include Amazon Elastic Container Registry \cite{ecr}, Azure Container Registry \cite{acr}, and Docker Hub Container Registry \cite{dockerhub}. This feature makes the PVC Data Loader a powerful data import tool with virtually unlimited use cases.

\subsubsection{Parallelism}

The PVC Data Loader must be able to pull large amounts of data into multiple PVCs simultaneously without the painstaking process of manually performing each transfer. A Kubernetes StatefulSet \cite{statefulset} deploys an ordered series of single container pods \cite{pods}, which can be mounted to the same PVC or ones with the proper IDs. That is to say, Pod-1 will mount to PVC-1, Pod-2 will mount to PVC-2, etc. The ID of each pod can be extracted inside each container through its hostname. The PVC Data Loader can achieve pleasantly parallel data transfers as follows: 

1. A text file containing the names of all files to be pulled is uploaded to each PVC to be loaded with data. 

2. The number of names can be divided by the number of Pods/PVCs to split the files between each one, and a range of files to be pulled by each container is derived by multiplying the ID of the Pod with the number of files to be pulled. If 100 files are to be split between 10 PVCs, Pod-1 will have the range 1-10, Pod-2 will have the range 11-20, etc. 

3. The entry point of the StatefulSet Pods is modified to calculate this range and pull the files into each mounted PVC.

4. The PVC Data Loader is deployed, and each pod begins pulling the files using the range and the text file on the PVC.

By applying this general method to the protocol needed, large sets of data can be easily split between any number of PVCs, all without any direct interaction between the user and the Pods.

\subsection{Loading PVCs}

Genomics datasets will be loaded as one dataset per PVC that is connected to one producer contianer.  For example, the publishing of 5000 named genome datasets will reside in 5000 PVCs.  However, it should be noted that PVCs do not need to be the same size and can scale to the data size.

\subsection{Data Lake}

\begin{figure}[h!] 
\centering
 \includegraphics[width=0.8\textwidth]{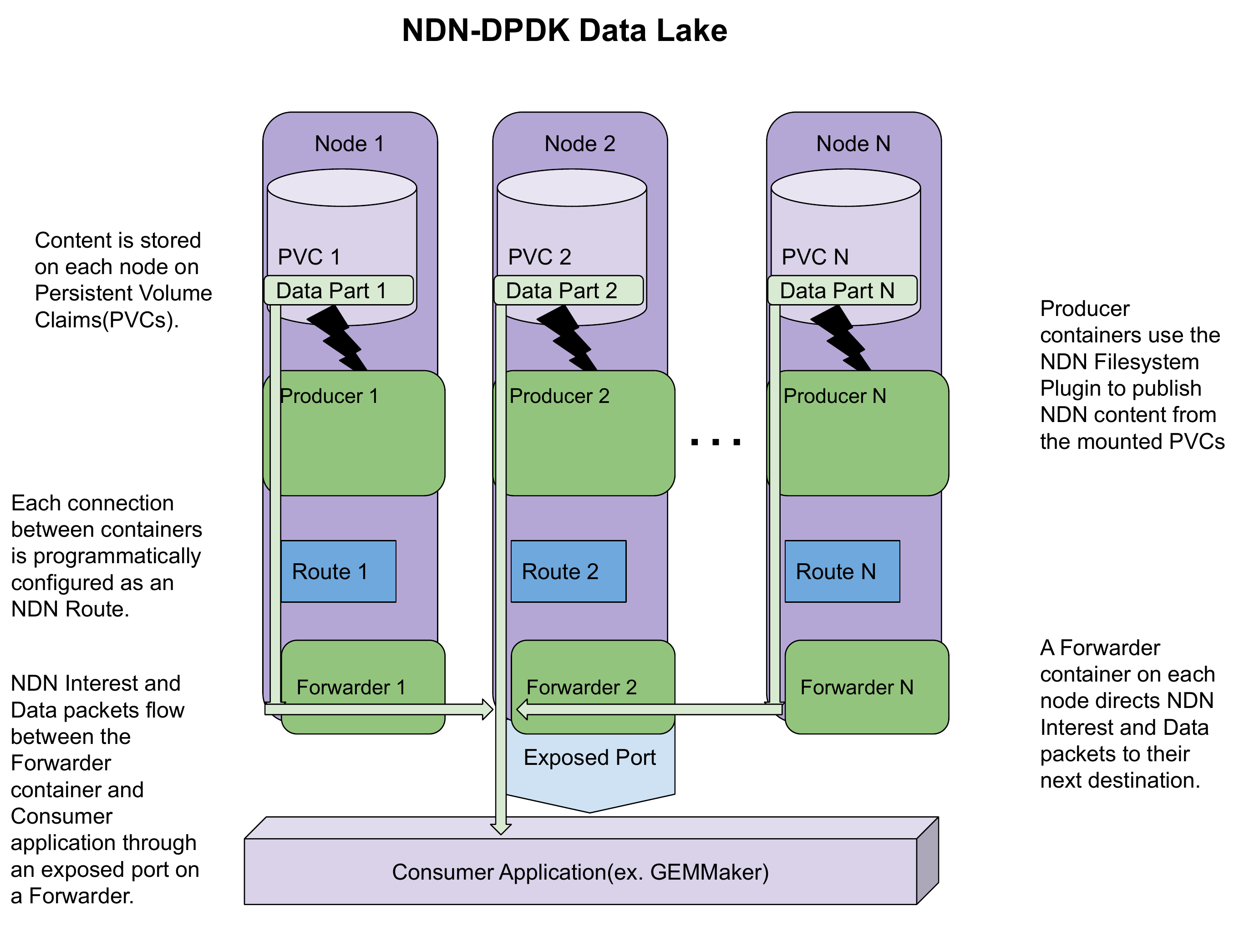}
 \caption{An overview of the current NDN Data Lake is provided using an example with at least three nodes. While all nodes have the ability to publish content through their own NodePorts, content is passed through a single NodePort to provide the information-centric experience to the consumer.}
\label{fig:data-lake}
\end{figure}

Multiple components are required for a robust deployment to create an NDN Data Lake. One of the first components of the design is the use of producer containers that act as producers to serve the data loaded into the PVCs. Next, the previously loaded PVCs are mounted to the producer containers to enable the producer to serve the stored data. After this, each producer container uses NDN routes to connect with a forwarder container. The forwarder container runs NDN DPDK and acts as a delivery interface between content consumers and producers. The connection between the producer and forwarder is made internally. This connection should be completed using MAC address over layer 2, but the current implementation requires layer 3 using cluster IPs. 

Due to the distributed nature of the data store, to allow data delivery regardless of which PVC the data is stored on, the forwarders can further be configured to communicate with each other inside the cluster. By default, work with Kubernetes does not allow communication requests from outside the cluster to access internal resources. NodePorts can map an external port on the node to a port on the hosted application to solve this. If forwarders in the data lake can communicate, every forwarder will not require a NodePort. Not every node in a Kubernetes cluster may be configured with NodePorts and an external IP. This design takes this into consideration by only requiring at least one node to have NodePorts. As long as at least one node has NodePorts configured, one or more data consumers outside the cluster will be able to access data stored on the Data Lake. The producer application runs as a systemd service on servers that have access to PVC data. Upon receiving the interest packets, the producer's job is to interpret the request and respond with data packets. The producer will perform the desired system call on the file, compose data packets, sign it and respond.

\section{Implementation}

\subsection{NDN-DPDK}

\paragraph{Build Steps}
When starting a Kubernetes deployment using kubevirt to deploy containers that require features like network interface control or anytime software needs to be added or updated, the first step is to create a VM. If utilizing the build scripts from the repository, "create-vm.sh" will assist with this step. The script is designed for deploying NDN-DPDK but can be modified to fit other purposes. The script downloads the Ubuntu 20.04 cloud image, sizes the VM to 5GB to optimize deployment resources, has the option to set a root password, creates a VM definition XML file, and starts the VM. Once the VM is created, the next step is to move the desired software in the VM.

This step is the recommended procedure for deploying NDN-DPDK; if using other software, this step can be skipped and software installed manually later before being deployed. Following the repository instructions, the next step is to acquire an up-to-date docker build of NDN-DPDK. Once the desired docker container is downloaded, the ID of the container is needed so that the "docker cp" command can be used to copy the files from the container to the local system. At this point, "virt-copy-in" can move these files into the VM. After the software binaries are moved, some additional configurations are needed before the VM can be used with Kubernetes. To accomplish these final steps before the software is installed, opening a console to the VM using "virsh" is required. The first step is to configure DHCP using "dhclient \&". With DHCP active, the next step is expanding the filesystem install to the full 5GB using "fdisk" and "resize2fs".

\begin{itemize}
\item Run the docker image that contains the binaries: docker run cbmckni/ndn-dpdk-builder:latest \&
\item Get the ID of the running container: ID=\$(docker container ls | grep cbmckni/ndn-dpdk-builder:latest | awk '{ print \$1 }')
\item Copy the files from the container to your local machine: docker cp \${ID}:/usr/local local
\item Copy the files from your local machine to the running VM: virt-copy-in -a ndn-dpdk.img ./local /usr
\item To get console to VM: virsh console ndndpdk
\begin{itemize}
\item Login with root and the password specified earlier
\end{itemize}
\item Start DHCP: dhclient \&
\item Resize partition: fdisk /dev/vda
\item Reboot for changes to take effect: reboot
\end{itemize}

With the basics complete, it is now time to install any desired software on the VM. Using NDN-DPDK, the "install-ndn-dpdk.sh" found on the repository can automate this, but otherwise, this can be done manually or with an alternative script. Once any needed software is installed and set up, the VM needs to be shutdown before being packaged into a container. With the VM fully configured, a docker container of the VM can be created using "docker build" and then pushed to a container store like DockerHub. Finally, once the container is created and pushed to a container store, the script "clean-up.sh" on the repository can be used to delete the VM and disk file.

\subsubsection{Deploy a Single NDN-DPDK VM on a Kubernetes Cluster}

Scripts and detailed documentation for this section are located at:


\paragraph{Installation} 

One of the first steps for a successful deployment is ensuring all necessary dependencies are satisfied. To deploy a VM on a Kubernetes cluster using kubectl, ensure that \textit{kubectl} is appropriately configured with the Kubernetes cluster of your choosing along with \textit{KubeVirt}. While kubectl will allow basic deployments, it is required to have \textit{virtctl} installed to gain access to a console. Console access allows a user to have a virtual terminal for the VM to run commands or change settings once started.

\paragraph{Deployment Steps} 

With the cluster environment configured, the next step is to deploy the VMs using kubevirt. Similar to working with containers, when working with VMs on Kubernetes, care must be taken to ensure that duplicates with the same name are not used. If a conflict is observed, the deployment script will then need to be modified or the existing VMs stopped.

\begin{itemize}
\item List running VMs: kubectl get vms
\item Edit the deployment script \textit{ndn-dpdk-vm.yaml} file if needed.
\item If there is already a VM with the name ndndpdkvm, change the name: and kubevirt.io/domain: fields to something unique.
\end{itemize}

With the deployment script verified not to cause a name conflict, the next step is to deploy, start, and access the desired VM. Deploying a VM pushes a configured VM and desired settings to a Kubernetes cluster. However, this VM deployment is not automatically started and still needs to be manually started. Once the VMs are started, console access is able to be facilitated using \textit{virtctl}. The root account credentials set when the VM was being built are used to access the terminal and gain access.

\begin{itemize}
\item Deploy the VM: kubectl create -f ndn-dpdk-vm.yaml
\item Start the VM: virtctl start <vm-name>
\item Get a console: virtctl console <vm-name>
\item Use the default password set for the VM with login root.
\end{itemize}

\paragraph{Clean Up} 
When any computation task is completed, "virtctl" and "kubectl" can be used to stop and clean up a deployment.
\begin{itemize}
\item Stop VM with: virtctl stop <vm-name>
\item Delete VM with: kubectl delete -f ndn-dpdk-vm.yaml
\end{itemize}

\subsection{Create and Load PVCs}

Scripts and detailed documentation for this section are located at:

\begin{itemize}
\item \textbf{https://github.com/cbmckni/pvc-data-loader}
\end{itemize}

PVC Data Loader is designed to help import data into Persistent Volume Claims. A scalable number of "loader" pods of any image will deploy to load data to newly created PVCs. Any containerized client can be used with this tool.

\subsubsection{Configuration}
The configuration file for the PVC loader is designed to be flexible in deployment, with edits done in the "helm/values.yaml" file. Helms is a deployment and automation tool that enables the easy installation of packages for any required dependencies. A helms file can also orchestrate the deployment of all the applications needed for a deployment.

\paragraph{Deployment Settings}
The first editable section in the configuration script is the deployment settings. By default, the StatefulSet PVC template will create Read-Write-Once(RWO) PVCs. If your use case requires Read-Write-Many(RWX) PVCs, create them manually(for now) before deploying. The options in this section include "Name", "Image", "Replicas", and "Arg". The "Name". variable is the name describing the data/location of the PVCs. Next, the "Image" variable is the image used to pull the data. After that is the "Replicas" variable, this setting defines the number of PVCs/primer pods to be created. Finally, "Arg" defines the entry point argument to either idle or start pulling data into "/workspace".

\paragraph{Resource Requests}
The next section in the configuration file is resource settings. This section allows modification and tuning of resources used for the deployed containers. The default configuration is left fairly conservative but can be scaled up or down depending on the needs of the deployment. Settings for CPU cores and RAM for both requests and limits can be defined here.

\paragraph{PVC Template Settings}
The final section in the PVC loader configuration file is PVC template settings. This section allows PVC settings to be defined for the deployment. The first variable is the "StorageClass" setting that defines the valid storage class to be used by the deployment. The second variable is the "Storage" setting that defines the storage per pod.

\subsubsection{Deployment}
With the configuration file refined to a particular deployment, the next step is to deploy the PVC Data Loader containers. This deployment file takes and applies the configuration to the Kubernetes spec.

\begin{itemize}
\item To deploy PVC Data Loader, run "helm install <DEPLOYMENT-NAME> helm/" from the "pvc-data-loader/" directory.
\end{itemize}

\subsubsection{Deletion}
Once any data loading procedures are complete, the primer pods can be deleted safely. Deleting the pods will not remove the primed PVCs. To remove the PVCs, "kubectl" must be used to delete the PVCs specifically.
\begin{itemize}
\item Delete pods: helm uninstall <DEPLOYMENT-NAME>
\item Delete PVC: kubectl delete pvc <PVC>
\end{itemize}

\subsection{Deploy Testbed}

Scripts and detailed documentation for this section are located at:

\begin{itemize}
\item \textbf{https://github.com/cbmckni/ndn-dpdk-k8s/tree/master/testbed}
\end{itemize}

\subsubsection{Installation}
One of the first steps for a successful testbed deployment is ensuring all necessary dependencies are satisfied. To deploy the outlined testbed on a Kubernetes cluster using kubectl, ensure that \textit{kubectl} is appropriately configured with the Kubernetes cluster along with \textit{KubeVirt}. The testbed deployment instructions also require that \textit{virtctl} is installed. 

\subsubsection{Deployment}
With the more restrictive deployment requirements than some Kubernetes clusters, a required step is to research the deployment environment to locate nodes with all required dependencies if all nodes do not. To create a list of compatible nodes on the intended cluster, save a list of names to a simple text file. The file comprised of this list is then used with the testbed deployment script and the testbed's name.

After creating a text file with nodes to use, the next step is to deploy the testbed. This deployment is accomplished using the deployment script in the repository. Once the deployment script is run, "kubectl" can be used to verify a successful startup and "virtctl" to gain console access.

\begin{itemize}
\item To deploy the testbed: ./deploy.sh <node-list-file> <testbed-name> 
\item To view deployed VMs: kubectl get vms
\item To view services: kubectl get svc
\item To gain console access: virtctl console <vm-name>
\end{itemize}

While the outlined instructions will create a testbed on Kubernetes, it can only be accessed inside the cluster. To be practical for a larger Data Warehouse deployment, access to the resources and storage from the public Internet is necessary. To achieve this, one or more nodes in the cluster will require NodePorts. If this requirement can be satisfied, the VM can be exposed for ingress traffic. 

\begin{itemize}
\item To expose a node to the public Internet, run: virtctl expose virtualmachineinstance <vm-name> --name <service-name> --type NodePort --port 63636 --target-port 6363 --node-port 63636
\end{itemize}

\subsubsection{Clean Up}
Once a testbed has satisfied the intended use, a script in the repository is included to manage easily cleaning up the deployment environment. 
\begin{itemize}
\item To destroy a testbed: ./clean-up-testbed.sh <node-list-file> <testbed-name>
\end{itemize}

\subsection{Publish Data}
Once the PVCs are loaded as described above, they are then mounted to nodes which run the fileserver along with the NDN-DPDK forwarder. Upon receiving an interest, the fileserver serves this data directly from the mounted volumes. These publication names are flexible but as a recommendation may follow a naming scheme like "/genomics/data/SRA/9605/…".

\subsection{Pull Data}
With a testbed created and data published, NDNc \cite{ndnc} was used as a client 
to assess the data retrieval performance  
as seen by the end-users. First, data was retrieved from another node inside the same compute cluster but geographically separate location. This experiment simulates a user whose institution participates in a distributed private cloud. Both layer 2 and layer 3 tests were done to see the performance differences. Finally, data retrieval tests from external to the compute cluster were run. The client was run on Google Cloud Compute Platform for these tests to provide a reliable and consistent end-point for experiments. 


\section{Limitation}
Kubevirt used as a part of this project helps running virtual machines on the top of Kubernetes. Virtual machines are comparatively more resource-intensive than directly running containerization technologies such as Docker on Kubernetes. This would eliminate the overhead of virtual machines and would make the deployment much faster and lightweighted.
\section{Conclusion}

This technical paper examines the growing need for a platform to assist with the growing availability of genomics data. We discussed NDN, a novel but well-researched future Internet architecture that can address these challenges at the network layer. We also examined Kubernetes, which provides a scalable way to deploy in the cloud. To facilitate the use of these technologies, we presented a design and provided instructions to allow a user unfamiliar with these technologies to deploy a Data Lake platform successfully. Finally, we examined ways that future research can be used to future improve the design for even greater flexibility and performance. 

\bibliographystyle{plain}
\bibliography{references.bib}

\begin{thebibliography}{10}

\bibitem{ndnc}
sandie-ndn/ndnc at master · cmscaltech/sandie-ndn.
\newblock [Online; accessed 14. Oct. 2022].

\bibitem{FAIR}
{GO FAIR initiative: Make your data {\&} services FAIR}, Jun 2020.
\newblock [Online; accessed 11. Feb. 2022].

\bibitem{dpdkwebsite}
{Home - DPDK}, Nov 2021.
\newblock [Online; accessed 14. Feb. 2022].

\bibitem{volumeclaimtemplates}
{StatefulSets}, Jun 2021.
\newblock [Online; accessed 11. Feb. 2022].

\bibitem{statefulset}
{StatefulSets}, Jun 2021.
\newblock [Online; accessed 11. Feb. 2022].

\bibitem{awscli}
{amazon/aws-cli - Docker Image {$\vert$} Docker Hub}, Feb 2022.
\newblock [Online; accessed 11. Feb. 2022].

\bibitem{acr}
{Azure Container Registry {$\vert$} Microsoft Azure}, Feb 2022.
\newblock [Online; accessed 11. Feb. 2022].

\bibitem{clusternetworking}
{Cluster Networking}, Jan 2022.
\newblock [Online; accessed 11. Feb. 2022].

\bibitem{dockerhub}
{Docker Hub Container Image Library {$\vert$} App Containerization}, Feb 2022.
\newblock [Online; accessed 11. Feb. 2022].

\bibitem{ecr}
{Fully Managed Container Registry {\textendash} Amazon Elastic Container
  Registry {\textendash} Amazon Web Services}, Feb 2022.
\newblock [Online; accessed 11. Feb. 2022].

\bibitem{services1}
{Kubernetes Networking 101 - NGINX}, Jan 2022.
\newblock [Online; accessed 11. Feb. 2022].

\bibitem{ndnwebsite}
{Named Data Networking (NDN)}, Feb 2022.
\newblock [Online; accessed 11. Feb. 2022].

\bibitem{osimodel}
Open systems interconnection model, Feb 2022.
\newblock [Online; accessed 15 Feb. 2022].

\bibitem{pvcs}
{Persistent Volumes}, Jan 2022.
\newblock [Online; accessed 11. Feb. 2022].

\bibitem{accessmodes}
{Persistent Volumes}, Jan 2022.
\newblock [Online; accessed 11. Feb. 2022].

\bibitem{pods}
{Pods}, Feb 2022.
\newblock [Online; accessed 11. Feb. 2022].

\bibitem{Kubernetesio}
{Production-Grade Container Orchestration}, Feb 2022.
\newblock [Online; accessed 11. Feb. 2022].

\bibitem{EKS}
{Provide external access to Kubernetes services in Amazon EKS}, Feb 2022.
\newblock [Online; accessed 11. Feb. 2022].

\bibitem{k8sservices}
{Service}, Feb 2022.
\newblock [Online; accessed 11. Feb. 2022].

\bibitem{k8sservicestypes}
{Service}, Feb 2022.
\newblock [Online; accessed 11. Feb. 2022].

\bibitem{storageclasses}
{Storage Classes}, Jan 2022.
\newblock [Online; accessed 11. Feb. 2022].

\bibitem{micro}
{What are Microservices? {$\vert$} AWS}, Feb 2022.
\newblock [Online; accessed 11. Feb. 2022].

\bibitem{lake-aws}
{What is a data lake?}, Feb 2022.
\newblock [Online; accessed 11. Feb. 2022].

\bibitem{lake-google}
{What is a Data Lake? {$\vert$} Google Cloud}, Feb 2022.
\newblock [Online; accessed 11. Feb. 2022].

\bibitem{NFDdevguide}
Alexander Afanasyev, Junxiao Shi, Beichuan Zhang, Lixia Zhang, Ilya Moiseenko,
  Yingdi Yu, Wentao Shang, Yanbiao Li, Spyridon Mastorakis, Yi~Huang,
  Jerald~Paul Abraham, Eric Newberry, Steve DiBenedetto, Chengyu Fan, Christos
  Papadopoulos, Davide Pesavento, Giulio Grassi, Giovanni Pau, Hang Zhang, Tian
  Song, Haowei Yuan, Hila~Ben Abraham, Patrick Crowley, Syed~Obaid Amin, Vince
  Lehman, Muktadir Chowdhury, and Lan Wang.
\newblock Nfd developer's guide.
\newblock 2016.

\bibitem{afanasyev2018nfd}
Alexander Afanasyev, Junxiao Shi, Beichuan Zhang, Lixia Zhang, Ilya Moiseenko,
  Yingdi Yu, Wentao Shang, Yanbiao Li, Spyridon Mastorakis, Yi~Huang, et~al.
\newblock Nfd developer's guide.
\newblock {\em NDN-0021}, 2018.

\bibitem{pvcdataloader}
cbmckni.
\newblock {pvc-data-loader}, Feb 2022.
\newblock [Online; accessed 11. Feb. 2022].

\bibitem{ogle2021named}
Cameron Ogle, David Reddick, Coleman McKnight, Tyler Biggs, Rini Pauly,
  Stephen~P Ficklin, F~Alex Feltus, and Susmit Shannigrahi.
\newblock Named data networking for genomics data management and integrated
  workflows.
\newblock {\em Frontiers in big Data}, 4, 2021.

\bibitem{ndndpdkpaper}
Junxiao Shi, Davide Pesavento, and Lotfi Benmohamed.
\newblock Ndn-dpdk: Ndn forwarding at 100 gbps on commodity hardware.
\newblock In {\em Proceedings of the 7th ACM Conference on Information-Centric
  Networking}, ICN '20, page 30–40, New York, NY, USA, 2020. Association for
  Computing Machinery.

\bibitem{smarr2018pacific}
Larry Smarr, Camille Crittenden, Thomas DeFanti, John Graham, Dmitry Mishin,
  Richard Moore, Philip Papadopoulos, and Frank W{\"u}rthwein.
\newblock The pacific research platform: Making high-speed networking a reality
  for the scientist.
\newblock In {\em Proceedings of the Practice and Experience on Advanced
  Research Computing}, page~29. ACM, 2018.

\bibitem{NDNDPDK}
usnistgov.
\newblock {ndn-dpdk}, Feb 2022.
\newblock [Online; accessed 11. Feb. 2022].

\bibitem{zhang2014named}
Lixia Zhang, Alexander Afanasyev, Jeffrey Burke, Van Jacobson, Patrick Crowley,
  Christos Papadopoulos, Lan Wang, Beichuan Zhang, et~al.
\newblock Named data networking.
\newblock {\em ACM SIGCOMM Computer Communication Review}, 44(3):66--73, 2014.

\bibitem{lake-mircro}
ZoinerTejada.
\newblock {Data lakes - Azure Architecture Center}, Feb 2022.
\newblock [Online; accessed 11. Feb. 2022].

\end{thebibliography}

\end{document}